\documentclass{sig-alternate-10pt}

\usepackage{amssymb}
\usepackage{latexsym}
\usepackage{url}

\usepackage{graphicx}
\usepackage{epstopdf}
\usepackage{color}
\usepackage[font={small}]{caption}

\usepackage{enumitem}
\usepackage{hhline}

\usepackage{url}
\usepackage{cite}
\usepackage{xspace}

\usepackage{subcaption}
\usepackage{array}
\usepackage{multirow}

\newcommand{\hide}[1]{}

\begin{document}

\title{Lost in Space: Improving Inference of\\IPv4 Address Space Utilization}
\numberofauthors{1}
\author{
  \alignauthor Alberto Dainotti$^{\ast\dagger}$, Karyn
  Benson$^{\ast\dagger}$, Alistair King$^{\ast\dagger}$, kc
  claffy$^{\ast\dagger}$, Eduard Glatz$^{\ddagger}$, \\Xenofontas
  Dimitropoulos$^{\diamond\ddagger}$, Philipp Richter$^{\star}$, Alessandro
  Finamore$^{\circ}$, Alex C. Snoeren$^{\dagger}$\\[4pt]
  \affaddr{$^\ast$CAIDA, $^\dagger$UC San Diego, $^\ddagger$ETH Zurich, $^\diamond$FORTH, $^\star$TU Berlin, $^\circ$ Politecnico di Torino} \\
}

\maketitle

\begin{abstract}
One challenge in understanding the evolution of Internet infrastructure
is the lack of systematic mechanisms for monitoring the extent to which
allocated IP addresses are actually used.  In this paper we try to 
advance the science of inferring IPv4 address space utilization 
by analyzing and
correlating results obtained through different types of measurements. 
We have previously studied an approach based on passive measurements
that can reveal used portions of the address space unseen by active
approaches. In this paper, we study such passive approaches in detail,
extending our methodology to four different types of vantage points,
identifying traffic components that most significantly contribute to
discovering used IPv4 network blocks.  We then combine the results we
obtained through passive measurements together with data from active
measurement studies, as well as measurements from BGP and
additional datasets available to researchers. Through the analysis of
this large collection of heterogeneous datasets, we substantially
improve the state of
the art in terms of: (i) understanding the challenges and opportunities
in using passive and active techniques to study address utilization;
and (ii) knowledge of the utilization of the IPv4 space. 
\end{abstract}

\section{Introduction}

In 2012, APNIC and RIPE exhausted their IP address pools;
the other Regional Internet Registries (RIR) will likely
exhaust their pools soon~\cite{ipv4_report_huston}. 
Running out of the Internet address space has been
anticipated for decades, accompanied by intense
debates over address management policy, IPv6 transition, 
and IPv4 address markets~\cite{huston2003ipv4,huston2007ipv4,
  huston2008changing}. However, only one project has invested
considerable effort in attempting to measure how many
allocated addresses are actually being visibily used
~\cite{heidemann-imc-2008}, where {\em used} was defined
as ``directly responding to an ICMP echo request''.
(Answer: less than 4\% of routed addresses.) 
In this study our objective is to build on this landmark work,
by refining the definitions of {\em used}, and 
extending the measurement methods to include other types of data,
including those that can reveal address usage not visible to ICMP. 

\begin{figure}
\includegraphics[width=0.45\textwidth]{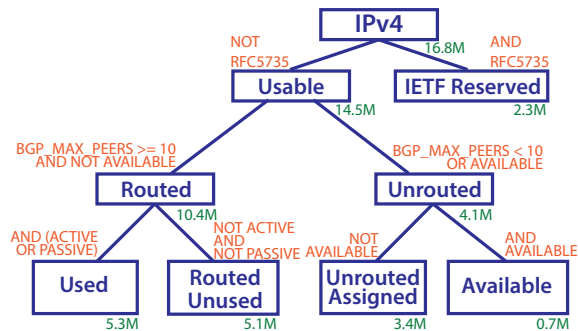}
\caption{IPv4 address space taxonomy. Nodes are labeled with the
  estimated /24 population of each category
  (Section~\ref{sec:census2013}) and the filter applied to arrive at
  the estimate (Sections \ref{sec:tree} through
  \ref{sec:buildcensus}).}
\label{fig:tree}
\end{figure}

Figure~\ref{fig:tree} 
(explained in more detail in \ref{subsec:data:alloc-bgp})
taxonomizes addresses as {\em IETF
  reserved}~\cite{rfc5735} and {\em usable}, the latter category
we further classify according to whether they are 
assigned to an organization, visible
to the global BGP interdomain routing system, 
and/or observably sending traffic. 
We have developed a methodology to measure and characterize IP 
address blocks per this taxonomy, and we use active probing and
passive traffic measurement to distinguish between the lower left 
leaves of the tree: {\em used} and {\em unused} address blocks.

Internet-wide
active probing poses at least four challenges: 1) measurement 
overhead, 2) potential violation of acceptable usage policies
triggering complaints or blacklisting of the measurement
infrastructure, 3) measurement bias due to operational filtering of
scanning; and 4) inability to scale for use in a future IPv6 census.
We recently showed~\cite{dainotti-ccr-2014} 
that passive measurements from darknets and an academic
network reveal an additional $\approx 450K$ active /24 address blocks not 
visibly active according to ICMP-based Internet census measurement. 
However, passive measurements introduce their own 
challenges, most notably the presence of traffic using spoofed 
source IP addresses, which can badly pollute estimates if not removed. 
In \cite{dainotti-ccr-2014}, we validated our methodology on two sources of traffic data available to us in 2012.

In this study, we first analyze the general applicability of 
passive measurements to survey Internet address usage. 
We extend the methodology introduced in~\cite{dainotti-ccr-2014},
the most important aspect of which is removing spoofed traffic,
to work with four different types of networks and
measurement data: (i) full packet traces from a large darknet; (ii)
netflow logs from a national academic network; (iii) sampled packet
traces from a large IXP; (iv) traffic classification logs from
residential customers of a European ISP.  We analyze how 
inferences of active address blocks can be influenced by 
characteristics specific to traffic observation vantage points,
such as traffic composition,  size of the monitored address
space, and duration and time of the measurement.
We found that our VPs were reasonably robust to variations
in these characteristics: we observe a substantial
fraction of address space at all VPs or when observing from
using smaller fractions of address spaces (where we could test that);
and each VP saw a consistent number of /24 blocks over
a two-year period.  

After gaining confidence in our methodology, we used 
seven passive and active measurement datasets 
collected from July 2013 through Oct 2013 to 
perform the first extended IPv4 Census using the taxonomy 
in Figure~\ref{fig:tree}.  
We compared our results to the state of the art represented by the ISI
census~\cite{heidemann-imc-2008} and found 718k previously undiscovered
used /24 blocks, an increase of 15.6\% over ISI.  We also
inferred used space from 98.9\% of the ASes announcing in BGP vs. 94.9\%
discovered by ISI, and obtained a visible increase in intra-AS
coverage (per-AS percentage of BGP-announced /24 blocks inferred as
used).

Our results show that only 5.3M /24 blocks are 
{\em used} (37\% of {\em usable} IPv4 space), 
and that 3.4M assigned /24 blocks are not even
visible in the global BGP routing system. 
We analyze how unused space is distributed across RIRs, 
countries, continents, and ASes. We inferred that
only 9.5\% of the legacy /24 blocks are {\em used} and that most 
{\em unused} address blocks are in the U.S.

Finally, we discuss how previous scientific
studies of Internet-related phenomena might change if they used this
extended dataset instead of other related data sets 
to estimate the address space of ASes or countries.

\begin{figure}[h]\centering
\includegraphics[width=1\linewidth, height=0.6\linewidth]{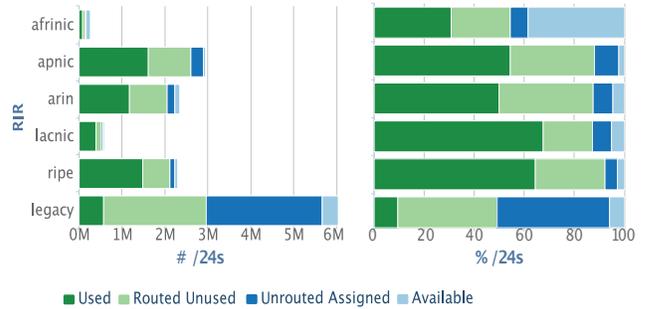}
\caption{Our final inferences classified by RIR-allocated (and legacy) 
	address space.  We identify legacy addresses per /8~\cite{iana-registry},
	but include some /8s that are presently administered by 
	RIRs. As is common knowledge, many legacy addresses are not used.}
\label{fig:per-rir-as-pct}
\end{figure}
Section~\ref{sec:related} and Section~\ref{sec:data}
describe related work and the datasets we use in our study.
Section~\ref{sec:tree} explains how we 
extract routed /24 address blocks from BGP and registry data. 
Section~\ref{sec:passive} provides a detailed evaluation of
our passive traffic methodology.  Section~\ref{sec:buildcensus} combines
passive and active measurement approaches to achieve 
unprecedented coverage in an IPv4 address space survey.
Section~\ref{sec:census2013} characterizes the
utilization of the address space and Section~\ref{sec:conclusion}
offers promising directions for applicability and extension of this work.
We will share all datasets we are allowed to release 
per the AUP of their owners. 

\section{Related work}
\label{sec:related}

Huston~\cite{ipv4_report_huston,huston2003ipv4,huston2007ipv4,
huston2008changing}
has provided a wealth of statistics and projections related to IPv4
address space allocation and announcement in global routing tables, although
he does not attempt to discern which routed addresses are actually {\em
used} (for any defintion).  
Meng {\it et al}.~\cite{Meng:2005} found that 90\% of IPv4
prefixes allocated from 1997-2004 appeared in the global routing system
within 75 days.

With respect to measurement to evaluate actual
address usage, the landmark work is USC's long-standing
effort \cite{heidemann-imc-2008}, based on 
comprehensive ICMP probing of the IPv4 space.  
Probing every routed IPv4 address over $\sim$30 days,
repeated multiple times between 2005 and 2007,
they observed only 3.6\% of allocated addresses responding
\cite{heidemann-imc-2008}.  In developing their
methodology, they compared ICMP and TCP probing to passive
traffic observation of USC addresses on USC's own campus network, 
finding 14\% more USC IP addresses visible to ICMP than to TCP, and 
28\% more USC IP addresses visible to passive traffic observation
than to either ICMP or TCP active probing. 
But each method observed some IP addresses missed by other methods.
We also found active and passive methods are able to
observe different subsets of addresses (Section~\ref{sec:buildcensus}),
but unlike~\cite{heidemann-imc-2008}, we use our passive monitors
to infer usage about the entire Internet instead of only
hosts internal to a network we monitor. 

In 2013 Durumeric {\it et al}.~\cite{zmap13} explored the system
challenges of active Internet-wide scanning in developing Zmap, a 
scanner that probes the entire IPv4 address space in under 45
minutes from a single machine. Accelerated scanning was also a goal
of an Internet Census illegally (and anonymously) performed in 2012 
from a botnet~\cite{carna_botnet}, although their methods were neither
well-documented nor validated \cite{caida-carna}. 

Others have also explored the use of passive data to estimate
specific usage characteristics of IPv4 addresses.
Zander {\it et al}.~\cite{zander13estimatingipv4} estimated the 
number of used IPv4 addresses by applying a capture-recapture method
for estimating population sizes on active and passive measurement logs
of IP addresses collected from sources such as web servers and spam
blacklists. In contrast, we also infer which IP addresses are {\em unused}.
Bartlett {\it et al}.~\cite{Bartlett:2007} found that passive 
traffic observation and active probing complemented each other for
the purpose of discovering active network services on campus.

Cai {\it et al.}~\cite{Cai:2010} explores (and undertake several)
potential applications of clustering active probes 
to infer address usage, including understanding
how efficiently individual address blocks are used,  
assessing the prevalence of dynamic address management, 
and distinguishing low-bitrate from broadband edge links.

\section{Datasets}
\label{sec:data}

\begin{table*}
\centering
\scriptsize
\begin{tabular}[t]{|c||l|l|l|}
\hline
Dataset & Source type & Data format & Period \\
\hline
\hline
UCSD-NT~\cite{ucsd_darknet}   & Traffic: Darknet     & full pkt traces  &  July 23 
	to August 25, 2013 \\
SWITCH~\cite{switch}   & Traffic: Live Academic Net.    & Netflow logs  & 
	July 23 to August 25, 2013 \\
IXP~\cite{large-european-ixp} & Traffic: IXP     & sFlow packet samples & July 8 to July 28, August 12 to September 8, 2013 \\
R-ISP~\cite{r-isp} & Traffic: Residential ISP & Tstat\cite{tstat-ieee} logs & from July 1 to September 31, 2013  \\
\hline
ISI~\cite{isi-dataset-used} & Active Probing: ICMP ping & logs & July 23 to August 25, 2013 \\
HTTP~\cite{sonar-http} & Active Probing: HTTP GET & logs & October 29, 2013 \\
ARK-TTL~\cite{topodataset} & Active Probing: traceroute & logs &  July to September, 2013 \\
\hline
BGP~\cite{routeviews, ripe-ris} & BGP announcements & RIBs & July to September, 2013 \\
Available Blocks~\cite{geoff-bogon} & IANA/RIRs & IP ranges & October 1, 2013 \\
NetAcuity Edge~\cite{netacuity} & IP Geolocation & IP ranges & July 2013  \\
prefix2AS~\cite{pfx2as} & BGP announcements & prefix to ASN & July 2013  \\
\hline
\end{tabular}
\caption{We infer used /24 blocks from passively collected traffic
(UCSD-NT, SWITCH, IXP, R-ISP) and active probing (ISI, HTTP, ARK-TTL).
The remaining datasets are used to infer both usable and routed prefixes,
or label prefixes according to geolocation and AS.
}
\label{tab:datasets}
\end{table*}

Table~\ref{tab:datasets} lists characteristics
of the datasets
-- collected between July and October 2013 --
from which
we extracted /24 blocks and inferred attributes.\footnote{We 
did not use reverse DNS PTR scans of the IPv4 space for
the same reasons articulated in \cite{heidemann-imc-2008},
namely that many active IP addresses lack DNS mappings, and
many unused IP addresses still have (obsolete) DNS mappings.}
\subsection{Address Allocation and BGP Data}
\label{subsec:data:alloc-bgp}

We analyzed BGP announcements captured by all collectors (24
collectors peering with 184 peers) of the Routeviews \cite{routeviews}
and RIPE RIS \cite{ripe-ris} projects. For each collector
we took all routing tables 
(dumped every 2 hours by
Routeviews and 8 hours by RIPE RIS) and 
built per-day statistics for each peer. 
For each /24 block, we computed the maximum number of peers 
that saw it reachable at any time within the full observation period of
92 days.

To determine which address blocks are available
for assignment, we used a dataset compiled by Geoff 
Huston \cite{geoff-bogon}, which merges
the extended delegation files from the 5 RIRs~\cite{delegated-arin,delegated-apnic,delegated-afrinic,delegated-lacnic,delegated-ripe} with IANA's published
registries \cite{iana-registry,iana-registry-special,iana-registry-ipv6-unicast,iana-registry-ipv6-special,iana-registry-as-numbers,iana-registry-recovered-address-space}.  We classified as {\em available} 
any /24 blocks falling in address ranges in this data set that were 
marked as either ``available'' 
(i.e., allocated to an RIR but not yet assigned to an LIR or organization)
or ``ianapool'' (i.e., IANA has not allocated it to an RIR).  
This data does not have LIR granularity, thus
	we considered any block allocated to an LIR as assigned (i.e.,
	not available).

We labeled as \textit{rfc5735} all /24 blocks within network
ranges reserved by IETF (private networks, multicast, etc.) \cite{rfc5735}.

\subsection{Passive Data-plane Measurements}

\label{sec:data:passive}

\begin{figure} 
\centering
\includegraphics[width=0.80\linewidth]{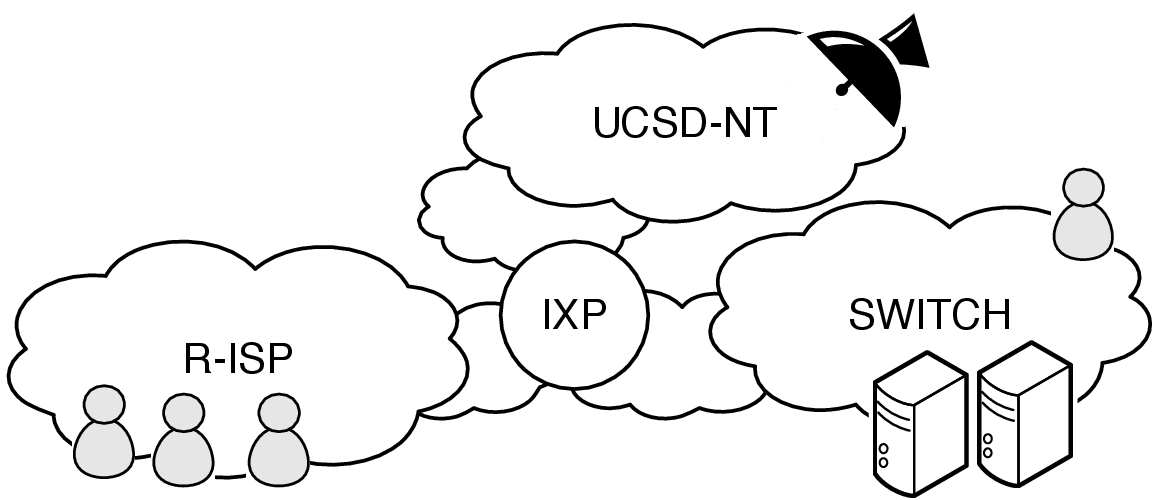} 
\caption{Our four traffic observation vantage points 
host different services and thus observe different workloads. 
They each have their own format for storing traffic data 
(sampled and unsampled packets, NetFlow records, flow-level traffic
classification logs), requiring substantial effort to 
curate them for use in our census (described in Section \ref{sec:passive})} 
\label{fig:vp_overview} 
\end{figure}

We apply our passive methodology for
inferring used /24 blocks to the following four vantage points (VP),
each of which retains traffic data in different formats 
(Figure \ref{fig:vp_overview}) and thus requires different approaches 
to filtering for use in a census (Section \ref{sec:passive}).

\textbf{SWITCH}. We collected unsampled NetFlow records 
from all the border routers of SWITCH, a national
academic backbone network serving 46 single-homed universities and
research institutes in Switzerland~\cite{switch}. The monitored
address range of SWITCH contains 2.2 million IP addresses, which
correspond to a continuous block slightly larger than a /11.

\textbf{R-ISP}. We collected per-flow logs from a vantage point monitoring
traffic of about 25,000 residential ADSL customers of a major
European ISP~\cite{r-isp}.  The VP is instrumented to run Tstat, an open
source passive traffic flow analyser~\cite{tstat-ieee}
that stores transport-level statistics of bidirectional flows,
and uses internal network knowledge to label flows as
inbound or outbound. 
\textbf{UCSD-NT}. We collected full packet traces 
from the /8 network telescope operated at the University of California
San Diego~\cite{ucsd_darknet}. 

\textbf{IXP}. Our fourth VP is a large European IXP interconnecting more
than 490 networks, exchanging more than 400 PB monthly~\cite{large-european-ixp}. 
We have access to randomly sampled (1 out of 16K) packets,
capturing the first 128
bytes of each sampled Ethernet frame exchanged via the public switching
infrastructure of this IXP. 
A sample includes full Ethernet, network- and
transport-layer headers, along with a few payload bytes.

\subsection{Active Measurements}

\textbf{ISI}.  We used the ISI Internet Census dataset
\textit{it55w-20130723}
\cite{isi-dataset-used}, 
obtained by probing the routed IPv4 address space with ICMP echo requests
and retaining only those probes that received an ICMP echo reply 
from an address that matched the one probed 
(as recommended \cite{isi-target-source}).
Note that the ISI Census experiment was designed to report at a
/32 (host) rather than /24 (subnet) granularity, but we apply the
resulting data set to a /24 granularity analysis.

\textbf{HTTP}.  We extracted IP addresses from logs of 
Project Sonar's HTTP (TCP port 80) scan of the entire IPv4 address space 
on October 29, 2013 \cite{sonar-http}.  For each /24 block, we stored
how many IP addresses responded to an HTTP GET query from the scan.

\textbf{Ark-TTL}. We processed ICMP traceroutes performed by 
CAIDA's Archipelago to each /24 in the routed IPv4 address space 
between July and September 2013 \cite{topodataset}.
Specifically, in order to add a third type of active probing data,
we extracted the ICMP Time Exceeded replies sent by hops along the
traceroute path.

\subsection{Mapping to ASes and Countries}

To establish a mapping from /24 block to ASN, we merged
all CAIDA's Routeviews Prefix to AS~\cite{pfx2as} mappings files for July
2013. For each /24 in the IPv4 address space, we identified the set
of overlapping prefixes and chose the most specific. 
We found 116k /24s (out of more than 10M) that mapped to multiple ASNs
(due to multi-origin ASes and AS sets), which we omitted from
our per-AS computations (Sections \ref{sec:buildcensus} and \ref{sec:census2013}). 

We geolocated each /24 block using Digital Element's NetAcuity Edge~\cite{netacuity} database from 6 July 2013. For each /24, we 
identified the unique set of country codes to which
overlapping blocks map.
We found 27k /24s (out of more than 14M) that map to multiple countries,
which we excluded from the geographic visualization in
Section~\ref{sec:census2013}.

\section{Un/Routed and Un/Assigned Space}
\label{sec:tree}
\textit{Which address blocks can we consider globally routed?
Of the unrouted space, which is assigned vs.~available?}

To distinguish legitimately routed address blocks from 
those that appear routed due to misconfigurations or hijacking,
we consider a /24 block as routed only if covered by a prefix
visible by at least 10 BGP peers.  RIPE recommends this
threshold \cite{ripebgpfiltering}, which we believe is 
reasonable since it removed 99.93\% of the /24 blocks we previously
determined are {\em available} or reserved by IETF
(Section~\ref{subsec:data:alloc-bgp}
and Figure~\ref{fig:tree}) and thus
could not be legitimately routed via BGP.
Applying this threshold, also excludes another 4.1M /24 blocks
that we would have otherwise labeled (likely incorrectly) as routed. 
We also filter out any other /24 blocks  
known to be in the \textit{available} category
(i.e., not assigned), even if observed by more than 10 peers.
Our filtering yields 10.4M {\em routed} /24 blocks that we must
further classify as {\em used} or {\em unused}.

Of the 4.1M {\em unrouted} /24 blocks (those we cannot observe in BGP),
we know that .7M are {\em available} and thus unassigned,
which leaves 3.4M /24 blocks that are assigned to organizations
(many of whom announce other IPv4 address space) 
and yet not routed.  
In other words, {\bf $\approx$53 /8's worth of address space 
are not used for the purpose of global BGP reachability.}

\section{Analysis of Passive Traffic}
\label{sec:passive}

\textit{How can we effectively extend our previous methodology
\cite{dainotti-ccr-2014} to different types of traffic?}

\begin{table*}
\centering
\scriptsize
\begin{minipage}{\linewidth}
\begin{tabular}{| l || c | c | r c || c | c | r  c |}
\hline
Vantage        &\multicolumn{4}{c||}{Original Traffic}          &\multicolumn{4}{c|}{After Applying Heuristics} \\
Point          & /24 blocks & Unrouted          & \multicolumn{2}{c||}{Dark}          & /24 blocks & Unrouted        & \multicolumn{2}{c|}{Dark}    \\\hline
UCSD-NT           &10,884,504  & 1,284,219 (31.6\%)& D-SWITCH:& 4,553 (90.9\%)& 3,152,067  & 2,123 (0.05\%)  & D-SWITCH:& 2 (0.04\%)          \\\hline
SWITCH         & 4,679,233  &    35,585 (0.69\%)& UCSD-NT:& 429 (0.68\%)    & 3,599,558  & 178 (0.004\%) & UCSD-NT:& 0 (0.00\%) \\\hline
R-ISP\footnote{
  Tstat automatically discards TCP flows not completing the 3-way handshake.  Our heuristics only remove UDP flows.
}
              & 5,233,871  & 344,188 (8.5\%)    & UCSD-NT:& 7,287 (11.6\%)    & 3,797,544 & 271 (0.006\%)   & UCSD-NT:& 0 (0.00\%)\\\hline
IXP           & 14,461,947 & 4,068,232 (78.5\%) &  UCSD-NT:& 62,838 (100\%)  & 3,091,021 &  376 (0.009\%) & UCSD-NT:& 3 (0.004\%)  \\\hline

\end{tabular}
\caption{\label{tab:summary_vantage_points}
For each VP, we report the absolute number and percentage of all /24
blocks that are unrouted.
For the dark category (4th and 7th column), we use the /24 blocks of
SWITCH
that did not generate bidirectional flows (D-SWITCH) to evaluate UCSD-NT, and the addresses
monitored by UCSD-NT to evaluate all other VPs.
Applying our heuristics reduces the number of
unrouted /24 blocks and dark /24 blocks at all VPs.%
}
\end{minipage}
\end{table*}

We build on our initial methodology \cite{dainotti-ccr-2014}
that we used to analyze traffic captured in 2012 at 
two of the four VPs described in Section \ref{sec:data:passive} 
and Figure \ref{fig:vp_overview}
(SWITCH and UCSD-NT). We extended this method
to work with the fundamentally different types of traffic
collected at the other two VPs (R-ISP and IXP), 
specially how to filter out spoofed traffic
(Section~\ref{subsec:spoofing}). 
We then evaluated the impact on our inferences
of varying aspects of the vantage points:
traffic composition %
size of monitored address space, %
and duration and times
of measurement (Section~\ref{sec:sensitivity}).

\subsection{Removing spoofed traffic} 
\label{subsec:spoofing}

The main challenge in curating  traffic data for use in
a census is to remove spoofed traffic
from the data sets, since it can severely distort
estimates of address utilization.  
Since the R-ISP data retains bidirectional flow information and
is guaranteed to see both directions of every flow,
filtering out spoofed traffic is easy.
For the IXP, the sampled data collection and the frequently 
asymmetric traffic flow (i.e., only one direction of a flow 
may traverse the IXP) mean that we cannot use the obvious 
and most reliable technique to infer spoofed traffic
(i.e., failed TCP flow completion, variants of which we use
for R-ISP and SWITCH data).
Indeed, the IXP data sees only one packet for the vast majority of flows.
The IXP traffic data also introduces a new challenge: filtering out 
packets with potentially unused destination addresses (e.g., scanning packets).

Although each VP's data set requires its own technique,
we tune and validate each technique using the same
assumption: packets appearing to originate from [or destined to]
\textit{unrouted} blocks are potentially spoofed [or scanning] 
packets.
As an additional source of validation, we compare 
our results at the SWITCH, R-ISP, and IXP VPs against
network blocks that we know to be unused, i.e., the dark /24
blocks in the UCSD-NT address space \footnote{Some addresses within this
``darknet'' are actually used and their traffic 
is not collected.} (62,838 /24 blocks).
To validate our estimates of spoofed traffic
at the UCSD-NT VP, we use
the /24 blocks from SWITCH that we infer to be dark because they did
not generate a single bidirectional flow in the whole observation period
(5,003 /24 blocks). 
	We use these data only with UCSD-NT because their
	observation periods exactly match.

\subsubsection{SWITCH (academic network)}
\label{subsubsec:SWITCH} 

To filter spoofed traffic, we use the same 
heuristic we introduced in~\cite{dainotti-ccr-2014},
which extracts from Netflow records
bidirectional TCP flows with at least 5 packets 
and 80 bytes per packet on average.  We performed
a sensitivity analysis on these thresholds 
in~\cite{dainotti-ccr-2014}, and found that they 
diminish the probability that
the remote IP address is spoofed. Using this heuristic
leads us to infer as used
only 0.004\% and 0\% of the unrouted 
and the UCSD-NT /24 blocks, respectively
(Table~\ref{tab:summary_vantage_points}).

\subsubsection{R-ISP (residential ADSL ISP)}
\label{subsubsec:R-ISP} 
Unlike the other traffic data sources,
the R-ISP's use of Tstat automatically removes essentially
all spoofed traffic, since to be logged a TCP flow must complete
the 3-way handshake.  For UDP traffic, we extracted
only bidirectional flows initiated locally 
with at least 1 packet with payload transmitted in both directions.

\subsubsection{UCSD-NT: (a large darknet)}
\label{subsubsec:UCSD-NT}

In~\cite{dainotti-ccr-2014} we looked deeply into several
{\em spoofing events} to derive filters that would allow us to
filter such events from darknet traffic in general. 
Two phenomena that we found to be indicators of a 
spoofing event were: 
(i) spikes in the
numbers of both unrouted and overall /24 blocks per hour, and (ii) traffic
using the same ports and protocols with a high fraction of unrouted 
source /24 blocks.
We derived heuristics to filter the most recurring
spoofing behavior (Table~\ref{table:darknet_spoofing_types}), plus some
specific large spoofing events (grouped in the table as ``All Specific
Filters'').  Many types of spoofing observed in our 2012
study~\cite{dainotti-ccr-2014} were also present in 2013. 
In addition,
we added two filters: TCP packets with no flags set and UDP packets
without payload.
\begin{table}
\centering
\scriptsize
\begin{tabular}[htbp]{|p{.38\linewidth}||c|c|}%
\hline
Filter Type&Total /24s&Unrouted\\\hline\hline%
TTL$>$ 200 and not ICMP&10,588,879&1,278,027\\\hline%
Least signif. byte\newline src addr 0&45,382&7\\\hline%
Least signif. byte\newline src addr 255&444,346&6,691\\\hline%
Non-traditional Protocol&56,502&2,209\\\hline%
Same Src. and Dst. Addr.&96&0\\\hline%
No TCP Flags&3,449&638\\\hline%
UDP Without Payload&545&114\\\hline%
All Specific Filters&10,587,049&1,280,826\\\hline%
\end{tabular}
\caption{Types of spoofed traffic observed at UCSD-NT. Many types of spoofing
were also observed in our 2012 study~\cite{dainotti-ccr-2014}.
}
\label{table:darknet_spoofing_types}
\end{table}

After applying our filters, we observe more than 3 million /24 blocks.
Table~\ref{tab:summary_vantage_points} shows that our filtering
heuristics reduce traffic appearing to originate from unrouted or dark
networks to around 0.05\% (compared to 
31.6\% and 90.9\% unrouted
and dark blocks, respectively, before filtering). 

\subsubsection{IXP (large IXP)}
\label{subsubsec:ixp}

For the IXP, we considered only TCP traffic and discarded
TCP packets with the SYN flag set,
which reduced the number of observably used /24s from 14.4M to 5.7M /24s. 
We then used a heuristic that tries to filter out /24s observed
as used only due to spoofing noise.  This heuristics is 
based on the number of packets and average packet size 
from and to a given /24 block. 
The first heuristic imposes a trade-off between false positives and
false negatives: if we set the threshold high enough, we are more
likely to filter out /24s that contain only IP addresses being used
in spoofed source address packets.   But we will also lose many
legitimately used /24 blocks, especially since we only have
1:16K sampled packet data in to begin with.
The average packet size threshold complements the packet count
threshold by increasing the likelihood of retaining /24s that
are actually exchanging TCP payload. 

\begin{figure}[htbp]
  \centering
  \begin{subfigure}[t]{0.95\linewidth}
    \includegraphics[width=\linewidth]{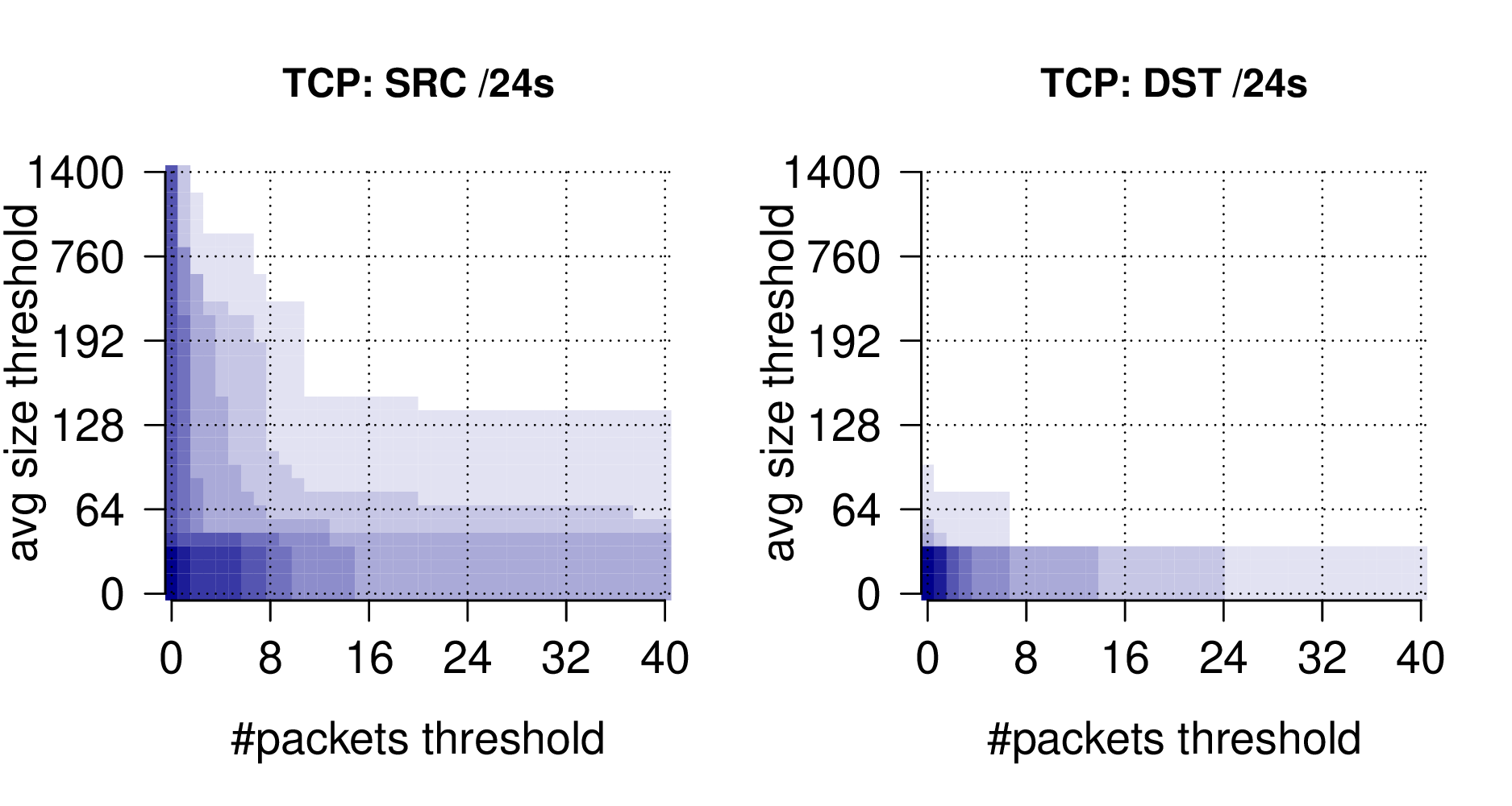}
    \caption{
      \label{fig:ixp_thresholding}
      Effect of thresholding: unrouted /24 blocks according to minimum
number of packets (x-axis) and minimum average packet size requirements (left).
Dark /24s inferred as used for DST addresses (right).
    }
  \end{subfigure}\quad
  \begin{subfigure}[t]{0.95\linewidth}
    \centering
    \includegraphics[width=\linewidth]{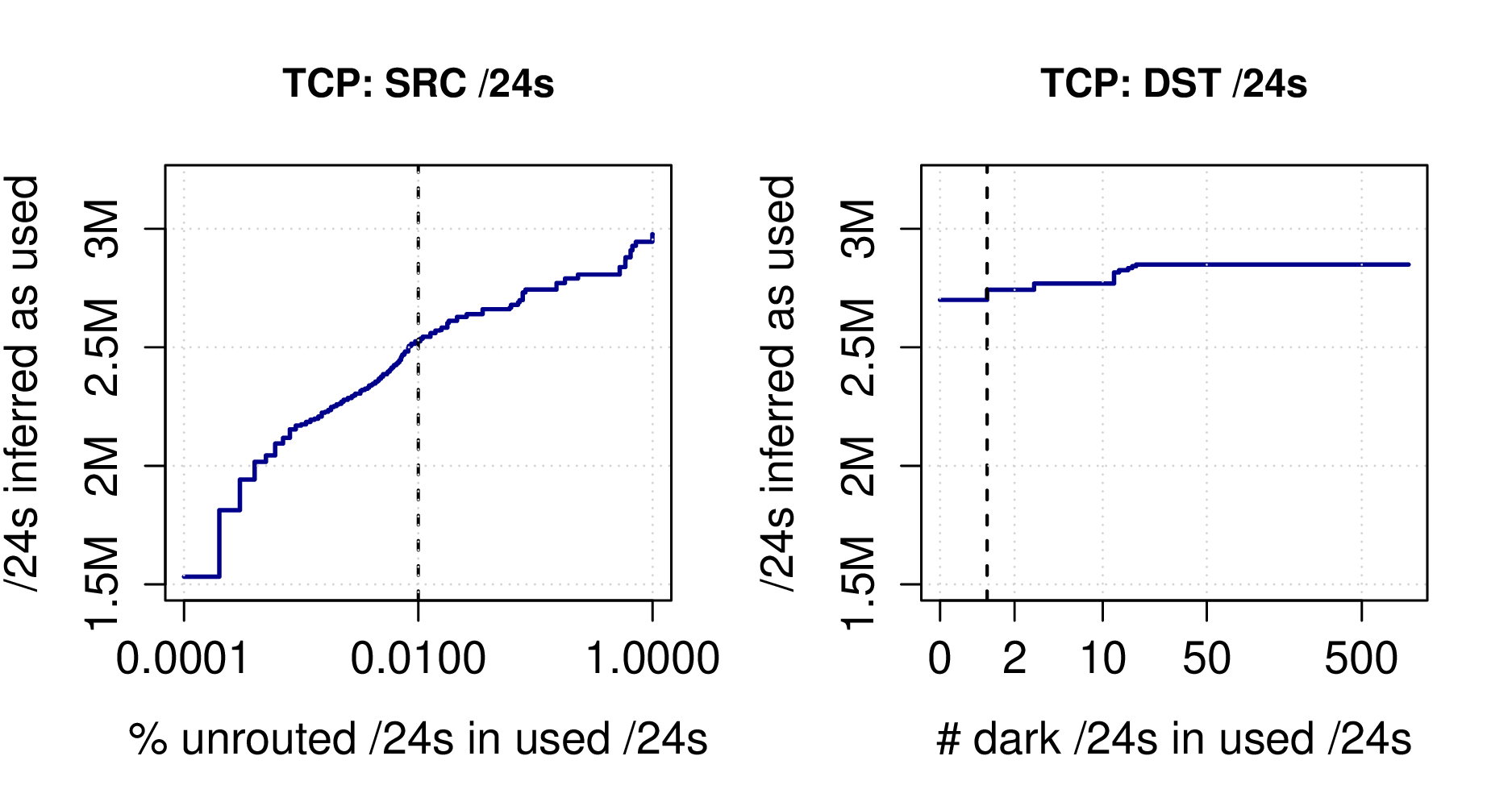}
    \caption{
      \label{fig:ixp_rocstyle} 
      Trade off between introduced error (unrouted /24s, dark /24s) and the number of /24s inferred as used.
    }
  \end{subfigure}
  \caption{
IXP: Selection of used /24s.
  }
\end{figure}

The left plot in Figure \ref{fig:ixp_thresholding} shows the number of
unrouted blocks that we inferred as used based on source addresses of 
the sampled packets (the darker the color the higher the number of 
unrouted blocks, log scale). 

Interestingly, we see almost no packets in the IXP data set 
destined to unrouted /24 blocks, perhaps because there are
no default routes advertised across BGP peering (vs.~transit)
sessions at the IXP, so only explicitly routed addresses will be 
observed as destinations.  We can still use dark but routed
destination addresses as indicators of scanning traffic; 
the right plot in Figure \ref{fig:ixp_thresholding} shows the number
of dark /24 blocks inferred as used when considering the 
destination addresses of packets. The average packet 
size is highly efficient at removing canning traffic.

To find an appropriate combination of thresholds, 
Figure \ref{fig:ixp_rocstyle} plots an ROC-like curve
that shows the number of /24 blocks inferred as used (y axis) against 
the percentage of unrouted (for source addresses) and 
number of dark (destination addresses) /24 blocks. 
For a given requirement (e.g., ``less than 0.1\%
unrouted''), we find the combination of thresholds (minimum 
number of packets and minimum average packet size) 
that results in the largest set of used /24 blocks.
To keep the error in our inference low, we consider a very conservative
threshold (as shown as the dashed vertical lines in Figure
\ref{fig:ixp_rocstyle}) and select used /24s from the SRC and DST addresses
independently. The resulting numbers are depicted in Table
\ref{tab:summary_vantage_points}.\footnote{Note that blocks inferred by
the source heuristic may also introduce 
dark /24s.} Our antispoofing approach
is efficient, reducing the number of unrouted and dark /24s dramatically,
even for sampled traffic. 
Nevertheless, we point out that the number of
used /24s directly depends on the thresholds applied and that the
false-negative rate increases with more conservative thresholds. Hence,
we likely miss used /24s with our current threshold selection.

We found similar behavior with UDP (as TCP) but we
needed to set thresholds more conservatively, 
particularly for average packet size.
We did not include UDP-based inferences in our final dataset,
since the additional gain in terms of /24s was not significant 
compared given our lower confidence in the inferences. 

\subsection{Effect of vantage points characteristics: traffic, network address segment, duration}
\label{sec:sensitivity}

After filtering spoofed traffic to the best of our ability, 
we analyzed the impact
of four characteristics specific to a given vantage point on the
number of /24s observed:  
traffic characteristics, size of address space monitored,
and duration or specific time of monitoring.  
We found that our VPs were reasonably robust to variations
in these characteristics, i.e, we observe a substantial
fraction of address space at all VPs or when observing from
smaller fractions of the address spaces (where we could test that),
and each VP saw a consistent number of /24 blocks over 
a two-year period.   

\subsubsection{Influential Traffic Components}

\label{sec:components}

\textit{How do traffic characteristics specific to a VP
influence its contribution to the inferences?} 

\begin{table}[!t]
\scriptsize
\begin{minipage}{\linewidth}
\centering
\begin{tabular}{|l |r@{$\,\,$}r |r | r |}%
\hline
R-ISP Traffic Class  & \multicolumn{2}{c|}{/24 Blocks} & Unique & Volume \\
\hline
\hline                                                                  %
P2P\footnote{eMule, ED2K, KAD, BitTorrent, PPLive, SopCast, TVAnts, and PPStream}
            &  3,172,439  &  (91.2\%)  &  610,438 & 34.1\% \\
Teredo      &  914,533    &  (26.3\%)  &  1,467   & 1.4\% \\
VoIP (RTP,RTCP)
            &  892,488    &  (25.7\%)  &  3,619   & 0.5\% \\
HTTP/HTTPS  &  234,586    &  (6.8\%)   &  20,274  & 57.7\% \\
Other\footnote{DNS, POP3, SMTP, IMAP4, XMPP, MSN, RTMP, SSH}
            &  196,503    &  (5.7\%)   &  62,406  & 1.9\% \\
Unknown\footnote{Flows unmatched by the classification engines.}
            &  2,691,300  &  (77.4\%)  &  115,869 & 4.5\%\\
\hline
\end{tabular}
\caption{At the R-ISP VP,
P2P traffic contributes almost 3.2M /24 blocks, 
including 610K unique.  HTTP/HTTPS is a smaller component,
despite accounting for 57.7\% of the volume.
}
\label{tab:isp-tstat-classification}
\end{minipage}
\end{table}

\begin{table}
\centering
\scriptsize
\begin{minipage}{\linewidth}
\begin{tabular}[htbp]{|l|r r|r|}
\hline
Darknet Traffic Class&\multicolumn{2}{c|}{/24 Blocks}&Unique%
\\\hline\hline
BitTorrent&
   2,210,257 &
   (70.2\%) &
   321,474
\\
Encrypted\footnote{Packets where entropy(payload) $\approx$ log$_2$ len(payload).}&
    1,349,578&
    (42.8\%)&
    34,290
\\
UDP 0x31&
    1,343,911&
    (42.7\%)&
    115,951
\\
Other P2P (eDonkey,QQLive)&%
    834,657&
    (26.5\%)&
    5,361
\\
Encapsulated IPv6 (Teredo,6to4)& %
    745,092&
    (23.7\%)&
    11,322
\\
Conficker&%
    604,877&
    (19.2\%)&
    61,836
\\
Backscatter&
    388,095&
    (12.3\%)&
    53,277
\\
Scanning (non-Conficker)\footnote{Meeting Bro's definition of a scanner: sent same protocol/port packets to at least 25 destinations in 5 minutes~\cite{bro-scan-parameters}.}&
    194,649&
    (6.2\%)&
    4,269
\\
Other&
    2,038,150&
    (64.7\%)&
    143,066
\\

\hline

\end{tabular}
\caption{At UCSD-NT,
BitTorrent traffic 
contributes the most /24 blocks,
instead of activities traditionally observed in darknets
(scanning, Conficker, backscatter).
}
\label{table:darknet_summary_phenomena}
\end{minipage}
\end{table}

Characterizing traffic at our VPs assists with two objectives: 
(i) highlighting how the VP contributes to the census;
and (ii) ensuring that traffic components specific to a VP
do not skew our findings or make them not generally applicable.
That is, to legitimately use passive traffic data for a census, 
we need to convince ourselves that a given VP is not
observing a special set of /24 blocks. %
Fortunately for our purpose, we found that the number of /24s 
we inferred did not vary dramatically across 
VPs with substantially different traffic compositions. 
(We could not analyze traffic composition from the IXP 
due to the sampled packet capture.)

\textbf{SWITCH.}
SWITCH hosts many popular services
that attract end users to the monitored address space,
including: a website hosting medical information, a SourceForge mirror,
PlanetLab nodes, university web pages, and mail servers.
The most popular service at SWITCH, the medical information
website, exchanged traffic with hosts in 1.8M /24 blocks; however,
all other IP addresses in SWITCH combined to capture 96.7\% of 
all 3.6M blocks observed by SWITCH.   
The top 100 services in SWITCH each observe over 70K /24 blocks, and
collectively contribute 91.2\% of the /24 blocks observed at this VP.
Compared to the UCSD-NT and R-ISP vantage points,
SWITCH's value as a VP depends more on these popular IP addresses.
If SWITCH did not host its top 1000 most popular IP addresses,
(i.e., the top services), it would observe only 69.9\% of the /24 blocks
it otherwise observes, compared to 89.7\% and 97.5\% at R-ISP and UCSD-NT
respectively.

\textbf{R-ISP.} 
Table~\ref{tab:isp-tstat-classification} aggregates the 
Tstat-identified traffic categories observed at R-ISP 
into five traffic components
accounting for 97\%
of /24 blocks observed at the ISP.  While HTTP and HTTPS 
account for 57.7\%
of the traffic volume, they contribute only 6.8\% of the /24 blocks
observed at the VP.
Instead, the largest source of /24 blocks comes from client-to-client
communication (e.g., P2P and VoIP).
P2P is a key contributor,
as 610k /24 blocks are only observable through P2P traffic.

\textbf{UCSD-NT.}
The non-uniform nature of darknet traffic is well-known 
~\cite{Wustrow:2010:IBR:1879141.1879149,cooke-dsn-2006},
but our 2012 study of UCSD-NT and a darknet of comparable size 
observed a similar number of /24 blocks 
in a 34 day study (3.14M vs 2.98M) \cite{dainotti-ccr-2014}.
Surprisingly, P2P also plays a key role at the UCSD-NT VP, where
we observe 2.2M /24 blocks (357k unique)
from traffic with a BitTorrent payload
(see Table~\ref{table:darknet_summary_phenomena}),
probably caused by index poisoning
attacks~\cite{liang-infocom-2006}. 
To a lesser extent, networks with end users are exposed
through malware-infected hosts (e.g., Conficker and scanning).
Alternatively, the backscatter traffic (a result of
a spoofed DoS attack) reveals networks likely hosting services.
Two classes (UDP 0x31\footnote{A 58-byte with 
payload mostly \texttt{0x00} and ninth byte set to
\texttt{0x31}.
The most common destination port (39455) is observed but unclassified
in~\cite{jczyz13v6darknet,rossow_ndss_14}.} and Encrypted) are 
of unknown purpose.

\subsubsection{Impact of Vantage Point Size}
\label{sec:size}

\textit{What is the effect of vantage point size?}

Analyzing the effect of vantage point size on the number of
/24 blocks observed is not straightforward due to
the non-uniform nature of the monitored address space.
Notwithstanding the extraordinary popularity of some IP addresses,
as well as non-uniform assignment of hosts within an
address subnet, we found an interesting correlation:
the median number of /24 blocks observed is roughly 
proportional to the log of the number of monitored IP addresses.
Consistent with this observation, the marginal utility of monitoring an 
additional IP address declines as the size of the vantage point 
increases, as expected. 

\subsubsection{Impact of Time}
\label{sec:duration}
\textit{How does the duration or time of collection affect the inference of which /24s are used?} 

\begin{figure}[htbp]
    \centering
    \includegraphics[width=0.75\linewidth]{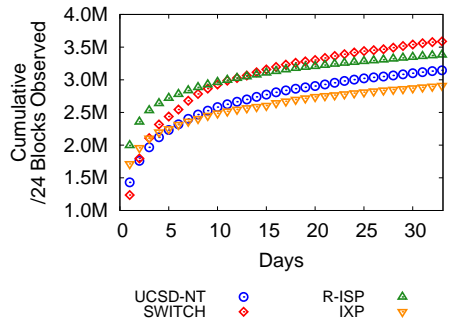}
    \caption{
      \label{fig:duration} The cumulative number of /24 blocks observed
      grows logarithmically at each vantage point.
    }
\end{figure}
\begin{figure}[htbp]
    \centering
    \includegraphics[width=0.75\linewidth]{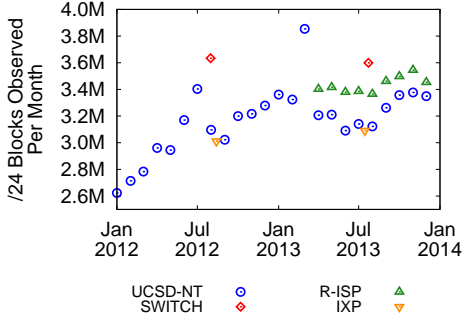}
    \caption{
      \label{fig:when}
      In our data, taken over two years, every VP observed at least 2.6M
      /24 blocks per month.
      SWITCH, R-ISP and the IXP show little variation in
      the number of /24 blocks captured by the VP per month, while
      UCSD-NT observes significant differences
      due to changes in IBR composition.
    }
\end{figure}

Figure~\ref{fig:duration} shows the logarithmic but varied growth of the
number of /24 blocks collected over time for our four VPs.  SWITCH, which
initially captures the fewest /24 blocks has the fastest growth rate;
while the R-ISP and IXP VPs capture more /24 blocks initially but they
grow
more slowly.  Other factors that can influence inferences are 
strong changes in traffic composition, e.g., flash events. 
Our traffic data sets all had low  (max 2\%) standard deviation in the 
number of /24 blocks observed per week, with no abnormal events observed.  

However, when observing measurements from a broader time frame,
we found evidence of flash events and changes in traffic.
For example, in August 2012 (the year before our study), SWITCH web sites 
hosting content about shark
protection experienced a sharp increase in visits (and thus
observed /24 blocks); the Discovery Channel's Shark
Week aired that month.
Figure~\ref{fig:when} shows per-month sample measurements 
using our methodology over a period of two years.
All vantage points except UCSD-NT observed a similar number of
/24 blocks per month.  At UCSD-NT, changes in IBR composition resulted in a
corresponding increase in visible /24 blocks.  Specifically, (i) in July
2012, there was an increase in BitTorrent traffic; (ii) in March 2013,
there was a large increase in the darknet's backscatter category, possibly
related to the DDoS attacks on Spamhaus~\cite{spamhaus-attack}.
Such events may bring additional data but note that for the last
two years each VP consistently observes over 3M /24s. 

\section{Combining Active and Passive}
\label{sec:buildcensus}

In this section, we first combine %
our seven datasets obtained from active and passive measurements to break
down the {\em routed} node in Figure \ref{fig:tree} into
{\em used} and {\em routed unused} categories
(we filtered all the datasets used in this section
to include only /24 blocks marked as routed according
to Section \ref{sec:tree}).
We then compare our results to the state of the art represented
by the ISI census (Section \ref{sec:improvement}).

\subsection{Active vs Passive}
\label{sec:a-vs-p}
\textit{
What are the respective contributions and limitations
of active and passive measurements?  
Are passive measurements from multiple VPs useful?}

The top half of Table \ref{tab:results} shows the number of /24 blocks
discovered by each active approach and their unique contribution.
The large number of /24 blocks found by ISI and HTTP,
and their distinct contributions
within the set of active measurements, are unsurprising
because we know that ICMP and TCP port 80 probing are
among the most effective active probing methods
that capture different but overlapping populations
\cite{heidemann-imc-2008, luckie-imc-2008}. 
More interesting is the 40k additional /24 blocks
that we obtain from the Ark dataset; 
we speculate that routers may be sending TTL exceeded 
packets using a source address from what they use in
ICMP echo responses.

\begin{table}[h!]
\centering
\small
\begin{tabular}{|c | c | c | c|}
\hline
\textbf{Dataset}		&  \textbf{\# /24s} & \textbf{\# Unique  /24s} & \textbf{\# Unique /24s}\\
		&    &  \textbf{within } & \textbf{among active} \\
		&    &  \textbf{same family} & \textbf{+ passive} \\
\hline
\hline
\textbf{Active} 		&		& & \\
\hline
ISI		&  4,589,213         & 1,319,283 & 398,334 \\
HTTP		&  3,161,064          & 189,831  & 76,189  \\
Ark-TTL		&  1,627,363          & 40,284   & 24,533 \\
\hline
\textit{Subtotal}		&  4,837,056	&	& \\
\hline
\hline
\textbf{Passive} 		&		& & \\
\hline
SWITCH		&  3,599,380         & 147,220 & 54,905 \\
UCSD-NT		&  3,149,944         & 61,443  & 24,134 \\
R-ISP		&  3,797,273         & 176,721 & 59,278 \\
IXP		&  3,090,645         & 195,328 & 55,155 \\
\hline
\textit{Subtotal}		&  4,468,096	&	& \\
\hline
\hline
\textbf{Total}		&  5,306,935	&	& \\
\hline
\end{tabular}
\caption{\small Each data set used to infer address space 
utilization offers a unique contribution.  
Unrouted /24 blocks are not represented here.
The third column is the number of /24s observed in the data set
that were not also observed in the (top) other active data sets 
or (bottom) other passive data sets;
the fourth column is the number /24s observed that were not
observed in any other data set. 
The final total is the number of /24s we infer as {\em used} 
(lower left node of tree in Figure~\ref{fig:tree}).}
\label{tab:results}
\end{table}

The bottom half of Table \ref{tab:results})
compares the contribution of our passive measurements.
The merged results from our four passive VPs 
do not entirely cover the set observed by
active measurements, missing about 840k /24 blocks.
However, the same data includes
470k /24 blocks not observed through active measurements,
demonstrating the value of combining active and passive datasets.
Each passive vantage point offers a
unique contribution, shown in the third and fourth columns of 
Table \ref{tab:results}, suggesting that these measurements 
are not exhaustive and that using more vantage points
would improve the coverage.
In particular, when we examine the portion of the address
space observed exclusively by passive approaches (470k /24 blocks,
not shown in the table),
we find that only 17\% of it was visible by all four
vantage points, while $\approx$ 41\% came from the sum of each unique
contribution (4th column in Table \ref{tab:results}).

Since 3 out of 4 vantage points are in Europe, we test
for the possibility of geographical bias in the passive measurements.
Table \ref{tab:geo-bias} shows the percent increase of
/24 blocks discovered by merged passive+active data 
vs. active measurements alone.  The larger increase
in European coverage vs.~other continents (middle column) 
is consistent with a slight bias from to the European 
vantage points, but on a per-continent basis the 
marginal increase spreads more easily across continents 
(right column, noting that the lower three continents 
have so much less address space that any increase will
be relatively large in percentage terms.)

\begin{table}
\centering
\scriptsize
\begin{tabular}{ l | c | c }
	&	\% of newly	&	per-continent \\
	&	discovered	&	\% increase \\
	&	/24 blocks	&	of /24 blocks \\
\hline
Europe		&32.44\%	&	11.11\% \\
North America	&26.54\%	&	9.08\% \\
Asia	&	25.31\%		&	7.64\% \\
South America	&8.56\%		&	10.85\% \\
Africa	&	4.65\%		&	30.18\% \\
Oceania	&	4.33\%		&	29.24\% \\
\end{tabular}
\caption{Absence of significant geographical bias in passive vs active
measurements: of the number of /24 blocks discovered by passive
approaches and not seen by active ones, a slight larger portion 
geolocated to Europe (where 3 of our 4 passive VP are). But
on a per-continent basis (right colum), the increase is more 
even across continents (Southern continents have little address 
space so any increase will be relatively large in percentage terms.)}
\label{tab:geo-bias}
\end{table}

We also explored why so much space is discovered by the active 
but not the passive measurement in our data sets. 
Perhaps our heuristics to remove spoofed traffic
are too conservative and remove much legitimate traffic. 
Also, for IXP and SWITCH, we included only TCP traffic
which could have limited our view; curating UDP and other traffic 
(removing spoofed and scanning traffic) is future work.

\begin{table}[h]
\centering
\scriptsize
\begin{tabular}{c | c | c}
\# VPs		& \# ISI-special /24s & \# single-IP /24s \\
		&		& without ISI-special 	\\
\hline
0		& 94,266 & 58,132 \\
1		& 13,057 & 19,414 \\
2		& 9,674		& 19,115		\\
3		& 4,959		& 27,185		\\
4		& 2,465		& 13,091		\\
\end{tabular}
\caption{Most /24 blocks with only a single IP address ending 
in .0, .1, .255 are not observed by any of our 
passive measurements (first row and middle column).  
In contrast, if a /24 had only a single responding address
ending in another octet, it was more likely to be 
observed sending traffic (3rd column).  We conclude that
/24s represented in the middle column likely do not send traffic
to the public Internet.} 
\label{tab:isi-special}
\end{table}

Our results also reveal large fractions of IPv4 space 
visible only by active measurements that do not generate 
traffic on the public Internet, shedding doubt on whether
they are really ``used'' for the purpose of global BGP reachability.
In particular, we found most /24 blocks from the ISI dataset
with a single responding address whose last octet was 0, 1, or 255
(the ``isi\_special'' column) were not observed in our 
passive measurements.  Table \ref{tab:isi-special} shows the
distribution of the number of passive vantage points that saw 
such /24 blocks (2nd column), as well as all /24s in the
ISI data that had only a single non-special
responding IP address (3rd column).  Many of these blocks were
not visible to any of our vantage points during the observation periods 
(Table \ref{tab:isi-special}, 1st row), including the vast 
majority of the ``isi\_special'' /24 blocks.
The progression from /24 blocks observed by 1 to 4 VPs shows a rapid
decay for ``isi\_special'' blocks (middle column), while there is almost
no trend for /24s in the right column. We conclude that
/24s represented in the middle column likely do not send traffic
to the public Internet.

We manually investigated other cases of network blocks only visible to
active probing, identifying special cases that suggest absence of traffic
on the public Internet, including clusters of /24 blocks apparently used
as internal CDNs by large service providers. We plan a more thorough
investigation of this behavior as future work.

The last row of Table \ref{tab:results} shows the final number (5.3M)
of /24 blocks we infer as {\em used} combining our 7 active and passive
datasets (leftmost leaf in Figure \ref{fig:tree}).  We subtract this
from the total amount of BGP-routed space (10.4M) to arrive at an
estimate of {\bf 5.1M {\em routed unused} /24 blocks, an
impressive quantity of unused but BGP-reachable IPv4 space}.
\subsection{Coverage}
\label{sec:improvement}
\textit{What is the improvement of our combined approach to infer
utilization in the routed space with respect to the state of the art (ISI
census)?}

We consider the ISI Census~\cite{heidemann-imc-2008} to be the state of
the art in inferring address space utilization within the routed space.
Since
there is no ground truth available about which routed space is actually
utilized, %
we present our results in terms of additional IPv4 space
coverage we obtain when combining our 7 datasets (which include ISI).
We consider coverage at three different levels:
(i) the percentage of routed /24 blocks inferred as used (\textit{global
coverage});
(ii) the percentage of ASes announcing the /24 blocks inferred as used out of
the ASes that announce at least one BGP prefix (44628 ASes)
(\textit{AS-level coverage});
(iii) for each AS, the percentage of routed /24 blocks inferred as used
(\textit{intra-AS coverage}).
AS-level coverage is the only case in which we expect the upper
bound to approximate ground truth (i.e., it is reasonable to 
assume that an AS announcing prefixes on BGP uses at least one /24 block).

\begin{figure}[h]\centering
\includegraphics[width=1\linewidth, angle=0]{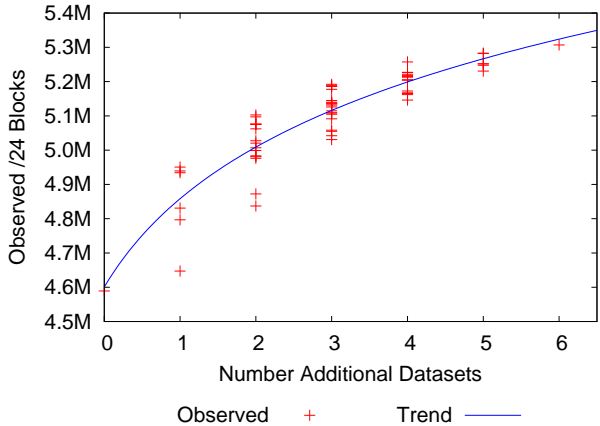}
\caption{\small 
We improve the global coverage of ISI (0 additional datasets) by
considering ISI + any number of our datasets.  The actual number of /24 blocks
observed are shown in red, and the seemingly logarithmic trend is shown in blue.}
\label{fig:global-coverage}
\end{figure}

We found 718k previously undiscovered used /24 blocks (difference between
last and 1st row of Table \ref{tab:results}), bringing global coverage from
44\% to 51\%.
Figure \ref{fig:global-coverage}
shows that
adding just a single dataset can greatly improve the
global coverage.  As we include our additional datasets, there is considerable
amount of overlap.
If we were to include additional measurements of used address space,
the actual number of /24 blocks would be highly dependent on the quality and
diversity of the datasets.
However, if we consider the logarithmic trend suggested by our observations,
increasing the number of additional datasets from 6 to 12 would result in
approximately 200k more /24 blocks.

Our AS-level coverage is 98.9\% versus 94.9\% found by ISI.
We manually analyzed whois and BGP data for the 489 ASes for which we
did not infer a single used /24 block.
We found that 37 ASes associated with U.S. military organizations
accounted for 79\% of the (17080) /24 blocks advertised by these
489 unobserved ASes.
We suspect such networks do
not transmit ICMP, TCP or UDP traffic over the public Internet
(but they may be tunneling traffic using, e.g., IPSEC,
which we did not capture in our passive measurements.)
The vast majority of the remaining ASes (399 out of 452) 
announce 10 or fewer /24 blocks. 

\begin{figure}[h]\centering
\includegraphics[width=1\linewidth, angle=0]{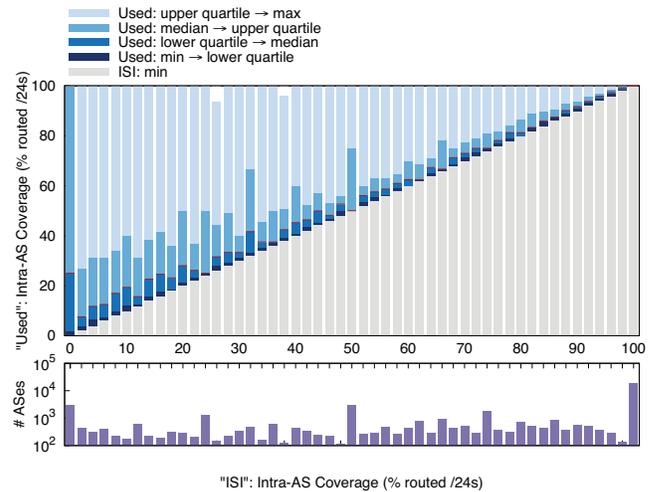}
\caption{\small Comparing the intra-AS coverage of our combined approach
(``Used'') against ISI's.  The graph is sorted by
increasing intra-AS coverage in ISI's data, with bins of 2\%. The bottom 
graph shows the number of ASes per bin.  In the top graph, the bottom grey
bar represents the minimum intra-AS coverage obtained by ISI for ASes
in the bin, whereas the remaining 4 (colored) bars refer to the intra-AS
coverage obtained by our combined approach (which includes ISI data).  
Each of
these 4 bars represents a quartile of the ASes in the bin.  For each
bar, its bottom and top show on the y axis, respectively, the lower and
upper bound of the coverage we obtain for ASes in that quartile 
(e.g., in the first bin, the bar from the median to the upper quartile shows
intra-AS coverage between 23\% and 100\%).}

\label{fig:as-coverage-over-isi}
\end{figure}

Figure \ref{fig:as-coverage-over-isi} 
shows the intra-AS coverage obtained with our combined approach as
a function of results obtained by ISI (the graph is sorted by
increasing ISI intra-AS coverage, with bins of 2\%). The bottom graph
shows the number of ASes per bin.  In the top graph, the bottom grey
bar represents the minimum intra-AS coverage obtained by ISI for the ASes
in the bin, whereas the remaining 4 (colored) bars refer to the intra-AS
coverage obtained by our combined approach (which includes ISI).  Each of
these 4 bars represents a quartile of the ASes in the bin.  For a given
bar, its bottom and top show on the $y$ axis, respectively, the lower and
upper bound of the coverage we obtain for such ASes. 
For example, in the first bin, the bar from the median to the upper 
quartile shows intra-AS coverage between 23\% and 100\%. 
The graph shows visible increments across the whole $x$ axis (decreasing
as ISI intra-AS coverage approaches 100\%). This result shows
that even for ASes which responded to ISI's pings ($x != 0$), our additional
datasets reveal new /24 blocks (i.e., ASes do not exhibit a uniform behavior
across their used subnets with respect to ICMP echo requests).
In most of the bins, for half of the ASes (i.e., two bottom quartiles) we
obtain a few percentage point increase. The two upper quartiles show 
more significant increments, e.g., up to $x=20$, for ASes in the 
upper quartile we see about 20\% more /24 blocks (at least). 
The first bin shows different behavior, with at least 25\% of 
ASes covered entirely by our method (although most of these
ASes announce only one /24).

SWITCH is the only AS for which we can derive better reference data
(rather than simply using the 100\% upper bound):
from 23 July to 25 August 2013, all 9,271 /24 blocks within SWITCH were
announced in BGP, but only 49\% of these blocks generated 
bidirectional flows.
Assuming these are the only used /24 blocks in SWITCH, 
 we should not infer an intra-AS coverage above 49\% for this AS
(instead of considering 100\% of the routed /24 blocks 
according to our definition of upper bound).
ISI's inferred 20.9\% intra-AS coverage for this
AS; our combined approach (without data from the SWITCH VP)
reached 33.1\%.
Still almost 16\% of the used blocks of the AS were not
discovered by our approach, showing space for further improvement.
However, for all other ASes we would include the SWITCH VP in our analysis,
potentially resulting in a higher intra-AS coverage.

\section{IPv4 Census 2013}
\label{sec:census2013}
\textit{How is used/unused and available space distributed across
  RIRs, ASes, countries and continents?  Which ASes or countries make
  the worst use of the space they have been assigned?
  Would previous scientific studies of Internet-related
  phenomena change if they used this dataset instead of 
	other related data sets?} 

Finally, we examine IPv4 address space utilization from the perspective of
our inferences. We emphasize that our inferences do not
provide complete coverage
of the used IPv4 address space, but it is the first dataset
which includes ASes and network blocks that do not reply to
ICMP probing. 
All our data is from approximately the same period (from July 2013
through Oct 2013). We assume that usage of the address space does not
change significantly within a period of 4 months.

\begin{figure}[tb]\centering
  \includegraphics[width=1\linewidth, angle=0]{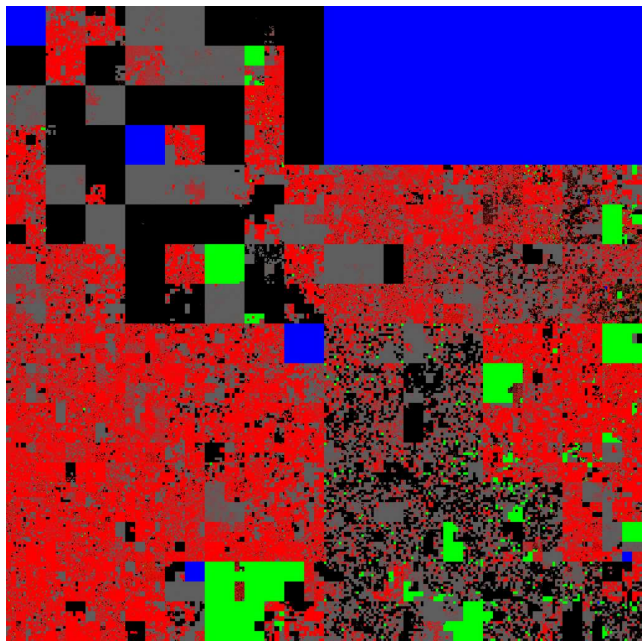}

\caption{Hilbert map visualization showing the utilization of the
  address space according to our taxonomy. The IPv4 address space is
  rendered in two dimensions using a space-filling continuous fractal
  Hilbert curve of order 12 ~\cite{hilbert-patent-1,xkcd}. Each pixel
  in the full-resolution image~\cite{supplemental} represents a /24
  block; \textit{red} indicates used blocks, \textit{green} unassigned blocks, and \textit{blue}
  RFC special blocks. Routed unused blocks are \textit{grey} and unrouted
  assigned \textit{black}.}
\label{fig:per-rir-absolute}
\end{figure}

Figure~\ref{fig:per-rir-absolute} illustrates a Hilbert map of IPv4
address space utilization based on our results, taxonomized in
Figure~\ref{fig:tree}.  The {\em IETF reserved} space accounts for
2.3M address blocks, or 13.7\% of the entire IPv4 address space (blue).
The remaining usable 14.5M address blocks consist of 5.3M (37\%) {\em used}
(red), 5.1M (35\%) {\em routed unused} (grey), 3.4M (23\%) {\em
  unrouted assigned} (black), and 0.7M (5\%) {\em available}
(green). Out of the 4.1M {\em unrouted} /24 blocks (Figure~\ref{fig:tree}), 3.4M /24 blocks are assigned to organizations
and yet not routed.
These numbers are striking and suggest revisiting the topic of IPv4
address depletion. First, an enormous amount of IPv4 address space is
assigned to organizations that do not even announce it on the BGP
plane (i.e., there is no need to perform inference through additional
active/passive measurements to sketch this phenomenon). In addition,
since we verified that several of these organizations announce on BGP
other address blocks they have been assigned, such number also
suggests that our inference of large unutilization of routed space is
realistic.

Figure~\ref{fig:per-rir-as-pct} classifed IPv4
addresses by their RIR region, or as {\em legacy} addresses
if they were allocated before the RIR system began. 
Legacy addresses were allocated by the central Internet
Registry prior to the RIRs primarily to military organizations and
large corporations such as IBM, AT\&T, Apple. Some of this space is
now administered by individual RIRs. We use the IANA IPv4 address
space registry~\cite{iana_legacy}, which marks legacy space and its
designation at a /8 granularity. In Figure~\ref{fig:per-rir-as-pct},
we observe that the set of legacy 
{\em routed unused} and {\em unrouted assigned} addresses
are similar in size (5.1M /24s) to 
the entire used address space (5.3M /24s). 42\% of the usable address blocks
are legacy; these blocks are more lightly utilized (9.5\% of the legacy) 
and include more unrouted assigned (45\% of the legacy) 
addresses than the RIRs (56\% and 7.7\% of the RIR address blocks, respectively).
ARIN, RIPE, APNIC, and LACNIC have 50\%, 65\%, 54\% and 68\%
of their address blocks used, respectively, in contrast to 
AFRINIC which has fewer of their blocks used 
(31\%) and many more available (38\%) address blocks
than other RIRs (6.7\% of other RIR addresses are available).

\begin{figure}[t]
\centering
\includegraphics[width=1\linewidth, angle=0]{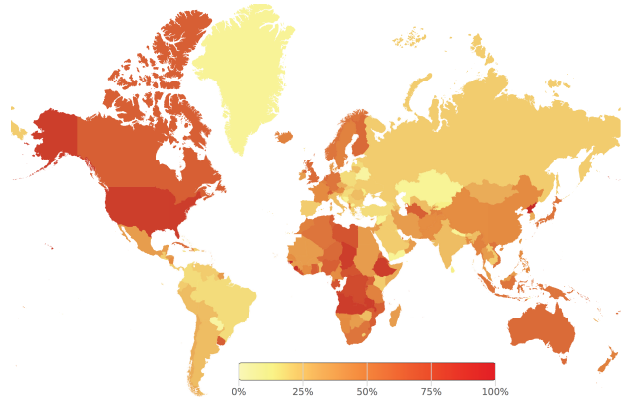}
\caption{Per-country percentage of unused space (\textit{routed unused} + \textit{unrouted assigned}) out of the assigned.
The U.S. is red in this map due to a 
few very large allocations heavily unutilized, while some African
countries are red because they use a very small fraction of their (also
small) assigned space.}
\label{fig:map-wasted}
\end{figure}

Table~\ref{tab:top-unused} lists the top-5 continents and
countries in {\em unused} and {\em unrouted assigned} 
/24s. 52.2\% of the unused
space and 72\% of unrouted assigned space is in North America, 
primarily in the U.S., where most legacy allocations (which
are only 9.5\% used, per Figure~\ref{fig:per-rir-as-pct}) were
made. Asia follows, with China owning 8.79\% and 5.7\% of the
global {\em routed unused} and {\em unrouted assigned} space, respectively, 
and then Europe. Other continents (South America, Oceania, and Africa) 
have between 0.93\% and 2.13\% of the global {\em unused} and 
{\em unrouted assigned}
space. We can also observe that the distribution of {\em unrouted assigned}
space is more skewed than {\em routed unused} space, because 45\% of
legacy space is {\em unrouted assigned}, while only 7.7\% of the
non-legacy space is {\em unrouted assigned}. Figure~\ref{fig:map-wasted}
visually illustrates the per-country ratio of {\em assigned unused} over
{\em assigned} space, suggesting which regions using space most
and least efficiently.  The U.S. is red in this map due to a 
few very large allocations, while some African countries are red because 
they use a very small fraction of their (also small) assigned space.

\begin{table}
\centering \scriptsize
\begin{tabular}{ l | c || l |c }
\hline
\multicolumn{4}{c}{Top Continents} \\
\hline
\multicolumn{2}{c||}{By Routed Unused /24s} & \multicolumn{2}{c}{By Unrouted Assigned /24s} \\
\hline
North America 	& 52.2\% 	& North America	& 72.0\% \\
Asia		& 22.3\%	& Asia		& 13.1\% \\
Europe		& 19.7\%	& Europe	& 12.1\% \\
South America	& 2.13\%	& Oceania	& 0.97\% \\
Oceania		& 1.92\%	& Africa	& 0.93\% \\
\hline \hline
\multicolumn{4}{c}{Top Countries} \\
\hline
\multicolumn{2}{c||}{By Routed Unused /24s} & \multicolumn{2}{c}{By Unrouted Assigned /24s} \\
\hline
USA		& 49.8\%	& USA		& 67.5\% \\
China		& 8.79\%	& China		& 5.70\% \\
Japan		& 6.22\%	& United Kingdom & 5.39\% \\
Germany		& 4.85\%	& Japan		& 4.21\% \\
South Korea	& 2.72\%	& Canada	& 3.73\% \\
\hline 
\end{tabular}
\caption{Top continents and countries in unused and unrouted assigned
  /24s. North America and USA have a large fraction of the assigned,
  but unused or unrouted address space.}
\label{tab:top-unused}
\end{table}

Table~\ref{tab:topas-unused} lists the top ASes by
{\em routed unused} /24s (we do not have per-AS data for {\em unrouted
assigned} space). The top ASes are the Department of Defense (DoD)
Network Information Center (NIC), followed by Level 3, HP, China
Telekom, and finally UUNET.

\begin{table}
\centering
\scriptsize
\begin{tabular}{ l | c }
\hline
\multicolumn{2}{c}{Top ASes in unused /24s } \\
AS Name \& Number & Routed Unused /24s (\%) \\
\hline
DoD NIC (721) 		& 190k (3.82\%) \\
Level 3 (3356) 		& 157k (3.16\%) \\
HP (71) 		& 126k (2.54\%) \\
China Telecom (4134) 	& 106k (2.13\%) \\
UUNET (701) 		& 105k (2.12\%) \\
\hline
\end{tabular}
\caption{Top ASes in routed unused /24s}
\label{tab:topas-unused}
\end{table}

\begin{figure*}[t]
\centering
\includegraphics[width=1\linewidth, angle=0]{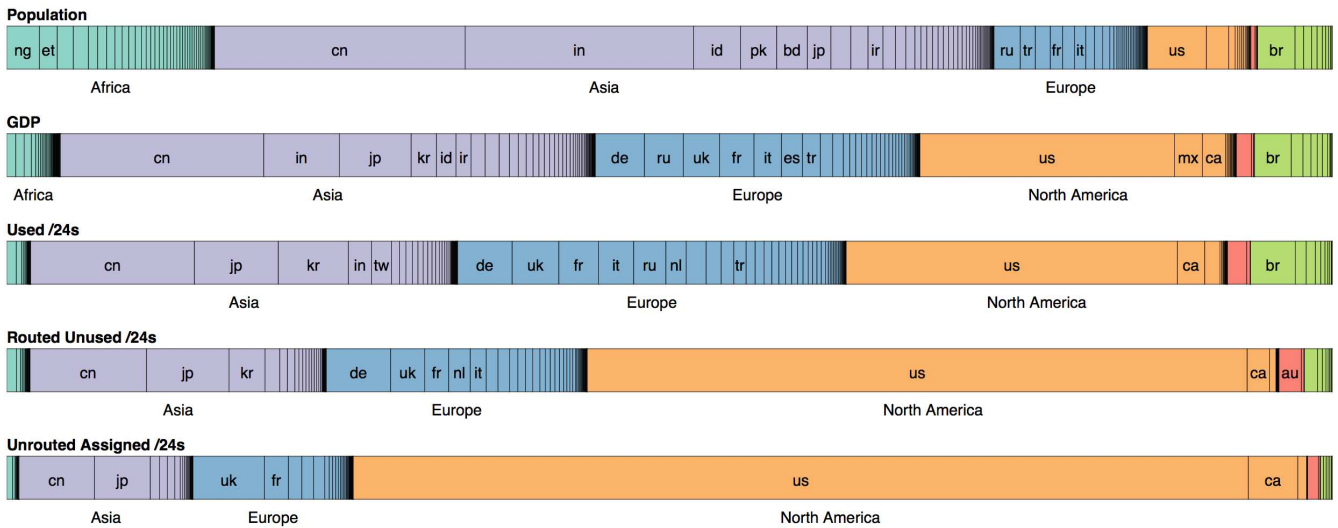}
\caption{Comparison of address space assigned to countries with
per-country population and GDP. The width of a country (and continent)
represents its relative size within a dataset. E.g., the top bar
shows the percentage that each country contributes to the global
population, with China (\textit{cn}) contributing the most (1.36B,
18.9\%). The
correlation between datasets can be observed by comparing bars. We
observe that there is not a strong correlation between population (top
bar) and number of {\em used} /24 blocks of a country; in large part due
to high usage by the USA. There is however, a strong correlation between
the GDP (2nd from top) and number of used /24 blocks of a country (3rd
bar). Not only does the USA dominate /24 block usage, it also represents
a significant portion of both the {\em routed unused} and {\em unrouted
assigned} bars, with 49.8\% and 67.5\% respectively. An interactive
version of this visualization is available at~\cite{supplemental}.}
\label{fig:inequality}
\end{figure*}

Figure~\ref{fig:inequality} compares address space assigned to
countries to per-country population~\cite{cia-population} and Gross
Domestic Product (GDP - we used ``purchasing power parity'' from CIA's
World Factbook~\cite{cia-gdp}). We observe notable disparities
between used /24s and population. For example, USA, Australia, UK,
Canada, and Germany have 25\%, 1.45\%, 3.52\%, 2.06\% and 4.11\% of
the used /24s, but only 4.44\%, 0.31\%, 0.89\%, 0.49\% and 1.13\%,
respectively, of the population. In contrast, African and Asian
countries have 16\% and 59\% of the population, but only 1.8\% and
32\%, respectively, of the used /24s. Nevertheless, the per-country
used /24s correlate much better with the distribution of GDP (0.960
correlation), than with population (0.517 correlation),
suggesting that economic inequalities could explain the differences in
the used /24s.

\begin{figure}[t]
\centering
\includegraphics[width=1\linewidth, angle=0]{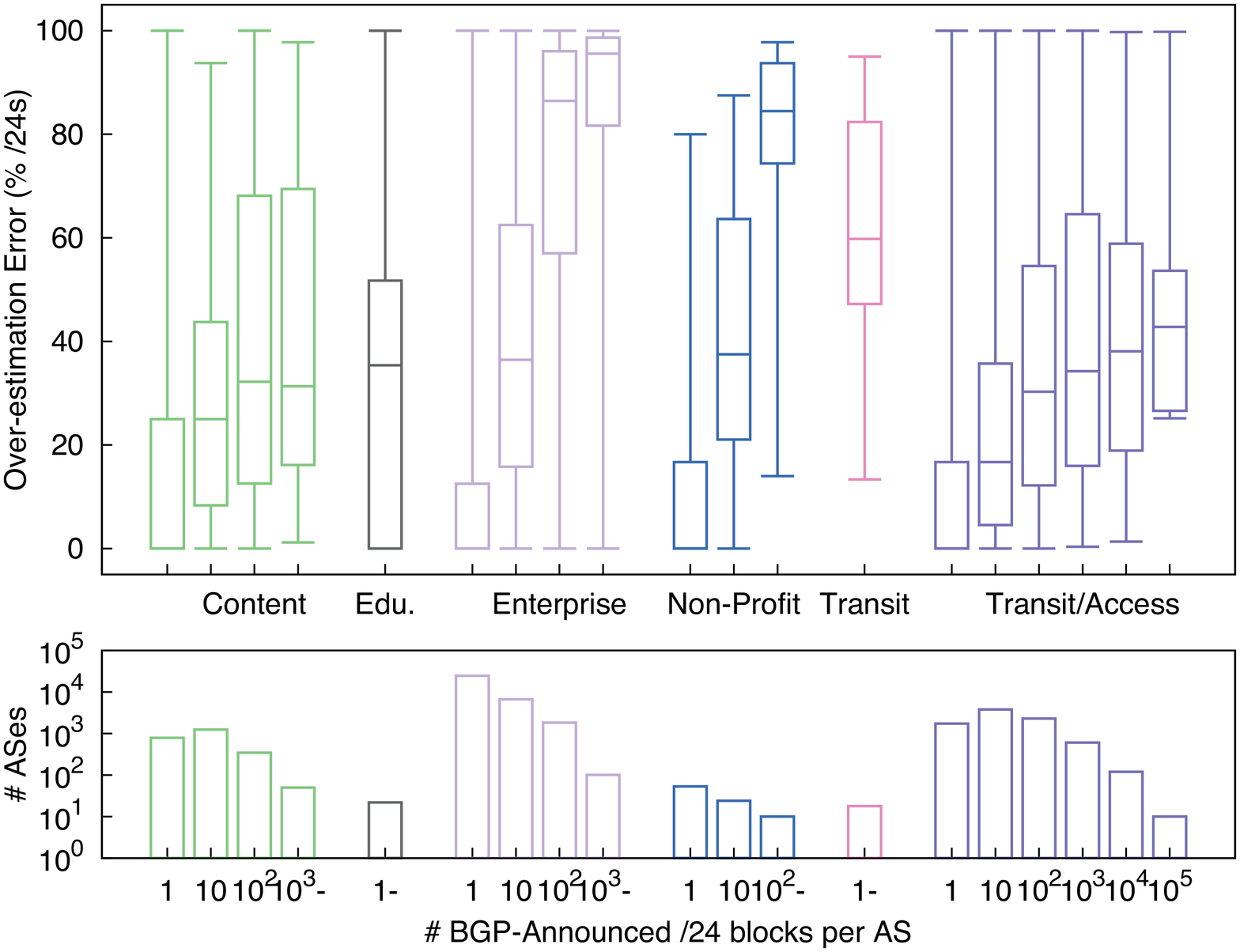}
\caption{Overestimation error (top graph) when using {\em routed address
space} instead of our census as a rough metric for AS size. ASes are
grouped according to the classification scheme proposed by Dhamdhere et
al.~\cite{amogh-ton11} and sorted by number of routed /24 blocks (the $x$
label indicates the minimum value in the bin).  The bottom graph shows
the number of ASes per bin. 
 Median overestimation error generally increases with the size of
the AS, perhaps due to large ASes under-utilizing their allocations.
Large Enterprise ASes ($>$1k /24s) result in the most dramatic
overestimation, with a median overestimation error of 96\%.}
\label{fig:bgp_error_1}
\end{figure}

\begin{figure}[t]
\centering
\includegraphics[width=1\linewidth, angle=0]{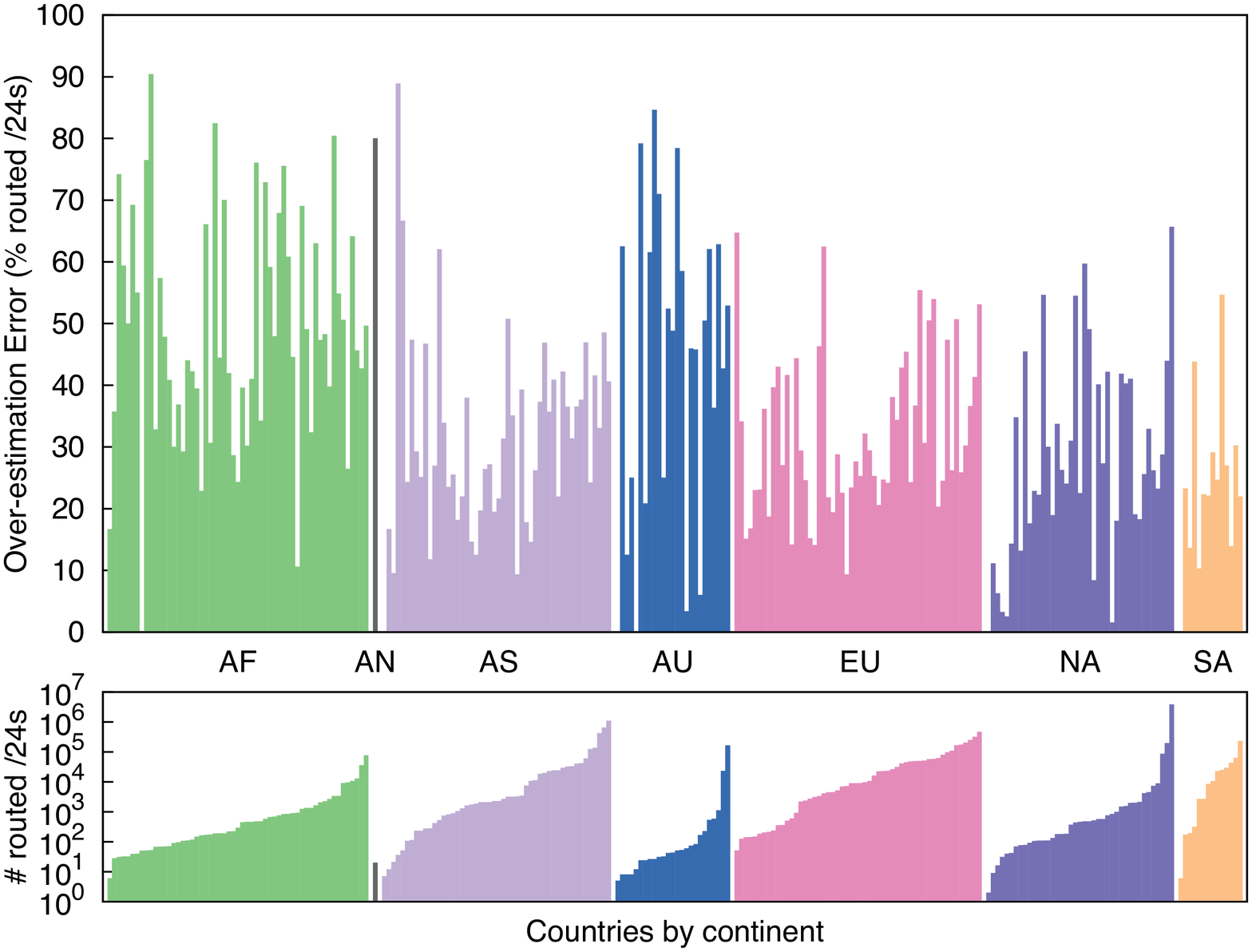}
\caption{Overestimation error when using {\em routed address space}
instead of our census as a rough metric for a country's footprint of
activity on the Internet.
Countries are grouped by continent and sorted by number of routed /24
blocks ($y$ value on bottom graph).
The top graph shows the overestimation error for
each country. As is also evident in
Figure~\ref{fig:map-wasted}, activity in African countries would be
significantly overestimated using {\em routed address space}. Most
importantly, there is no significant correlation between the
the number of per-country routed /24s and the resulting
overestimation error.}
\label{fig:bgp_error_2}
\end{figure}

This type of census dataset also has implications for a range
of scientific research of the Internet, most notably
projects that incorporate routed address space metrics 
into estimates of the size, degree, type, or maliciousness of
ASes~\cite{Tangmunarunkit:2001:SDD:1037107.1037108, 
lodhi-peeringdb,luckie13asrank, amogh-ton11,Konte:2012:RAM:2238896.2238912}.
More accurate metrics of address space usage 
could also potentially improve the accuracy of analysis
of (or prediction of likely future)
address blocks transfers in the 
grey market \cite{transfermarkets13}.
Figure \ref{fig:bgp_error_1} shows the overestimation error
one would make by using a canonical {\em BGP-routed address space}
metric to reflect how much address space an AS is actually
observably using, for five types of network providers
of various sizes. 
Figure \ref{fig:bgp_error_2} shows the overestimation error
when using the same ({\em BGP-routed address space}) to reflect
each country's Internet footprint. Both figures show that
there is no simple formula to translate between routed
address space and actually used address space -- the 
difference varies widely by AS. 

\section{Future Directions}
\label{sec:conclusion}

We presented a new methodology for performing an 
Internet-wide census of IPv4 address space utilization.
Among the many results presented, we find that only 
5.3M /24s address blocks are visibly {\em used} (37\% of 
{\em usable} /24 blocks), and that 3.4M assigned /24 blocks 
are not even visible in the global BGP routing system. 

In addition to the applications of census measurements 
that have been well articulated by~\cite{Cai:2010} 
(and summarized in Section~\ref{sec:related}),
there are many possible future directions for this work.
To improve the methodology, we would like to further improve 
our ability to infer spoofed traffic, and validate our 
inferences, perhaps by responding to darknet traffic.  
We would also like to investigate
the use of UDP or other protocol traffic at at R-ISP and IXP
vantage points, and analyze in more detail what addresses
are less visible to traffic measurement e.g., internal
CDNs or quiet networks.  As always, additional vantage
points and ground truth information from operators 
would help improve the integrity of the method.

For a periodic global Internet census that tracks changes
over time, we imagine a hybrid approach that first infers
active IP address blocks based on passive measurements from
one or more (live or dark) traffic vantage points, then
probes only those address blocks that cannot be confidently
inferred as active.   This approach could dramatically
improve coverage over state of the art methods, while
minimizing measurement overhead and potential irritation
of network operators with aggressive firewalls.  Finally,
the unscalability of active scanning to the IPv6 address
space was one motivation to explore our hybrid apporach,
but we do not know how well distributed passive traffic 
observation alone could effectively support a future IPv6 census.

\bibliographystyle{abbrv}
{\small
\setlength{\itemsep}{0pt}
\bibliography{imc2014}
}

\end{document}